\documentstyle[12pt]{article} \textwidth 14.5cm \textheight 
20cm
\begin{document} 
\newcommand{\add}{\addtocounter{eqncnt}{1}}
\newcounter{eqncnt}[section]
\newcommand{\al}{\mbox{$\alpha$}}
\newcommand{\med}[1]{g_{#1}} 
\newcommand{\meu}[1]{g^{#1}}
\newcommand{\be}{\begin{equation}\add} 
\newcommand{\ee}{\end{equation}}
\newcommand{\bea}{\begin{eqnarray}}
\renewcommand{\theequation}{\thesection.\theeqncnt} 
\newcommand{\eea}{\end{eqnarray}\add}
\newcommand{\und}{\underline}
\newcommand{\D}{\displaystyle}
\newcommand{\hs}{\hspace*{2in}}
\newcommand{\hqq}{\hfill\qquad}
\newcommand{\noin}{\noindent}
\setlength{\baselineskip}{20pt}

\title{A CLASSIFICATION OF SPHERICALLY SYMMETRIC SELF-SIMILAR DUST MODELS}
 
\author{B. J. Carr\\
Astronomy Unit, Queen Mary \& Westfield College,\\ Mile End Road, London E1 4NS, England \\
Yukawa Institute for Theoretical Physics, Kyoto University,\\ Kyoto 606-8502, Japan}
\maketitle
\begin{abstract}

We classify all spherically symmetric dust  
solutions of Einstein's equations which are self-similar in the sense that all dimensionless variables depend only upon $z\equiv r/t$.
We show that the equations can be reduced to a special case of the general perfect fluid models with equation of state $p=\alpha \mu$. The
most general dust solution can be written down explicitly and is described by two parameters. The first one (E) corresponds to the asymptotic energy at large $|z|$,
while the second one (D) specifies the value of
$z$ at the singularity which characterizes such models. The $E=D=0$ solution is just the flat Friedmann model. The 1-parameter family of solutions with $z>0$
and
$D=0$ are inhomogeneous cosmological models which expand from a Big Bang singularity at $t=0$ and are asymptotically Friedmann at large $z$; models with $E>0$ are
everywhere underdense relative to Friedmann and expand forever, while those with
$E<0$ are everywhere overdense and recollapse to a black hole containing another singularity. The black hole always has an apparent horizon but need 
not have an  event horizon. The $D=0$
solutions with
$z<0$ are just the time reverse of the
$z>0$ ones. The 2-parameter solutions with $D>0$ again represent inhomogeneous cosmological models but the Big Bang singularity is at $z=-1/D$, the Big Crunch
singularity is at
$z=+1/D$, and any particular solution necessarily spans both $z<0$ and $z>0$. While there is no static model in the dust case, all these solutions are
asymptotically ``quasi-static'' at large
$|z|$. As in the $D=0$ case, the ones with $E \ge 0$ expand or contract monotonically but the latter may now contain a naked singularity.
The ones with $E<0$
expand from or recollapse to a second singularity, the latter containing a black hole. The
$D<0$ models either collapse to a shell-crossing singularity and become unphysical or expand from such a state.  

\end{abstract}
\setcounter{equation}{0}
\section{Introduction}

Spherically symmetric self-similar solutions to Einstein's equations have the feature that every dimensionless variable is
a function of some dimensionless combination of the cosmic time coordinate $t$ and the comoving radial coordinate $r$. In the simplest situation, a similarity
solution is invariant under the transformation $r
\rightarrow at, t \rightarrow at$ for any constant $a$, so the similarity variable is $z=r/t$. Geometrically this corresponds to the existence of a homothetic Killing
vector. Such solutions have been the focus of much attention in General Relativity because the field equations simplify to ordinary
differential equations (Cahill \& Taub 1971). 

Even greater simplification is afforded if one focusses on the situation in which the
source of the gravitational field is pressureless dust since, in
this case, the solutions can often be expressed analytically and are just a
special subclass of the more general spherically symmetric
Tolman-Bondi solutions (Tolman 1934, Bondi 1947, Bonnor 1956). A number of people have studied such solutions:
for example, Gurovich (1967), Carr \& Hawking (1974), Dyer (1979), Wu (1981), Maharaj (1988), Carr \& Yahil (1990), Ori \& Piran (1990), 
Henriksen \& Patel (1991) and
Sintes (1996).  These papers are described in some detail by Krasinski (1997). As examples of applications of particular physical intererst, self-similar dust
solutions have also been used in studying naked singularities by Eardley \& Smarr (1984), Christodoulou (1984), Lake (1992), Joshi \&
Dwivedi (1993), Joshi \& Singh (1995), Dwivedi \& Joshi (1997) and in modelling cosmic voids by Tomita (1995, 1997a, 1997b)
and Carr \& Whinnett (1999). 

In this paper we will present a classification of spherically symmetric homothetic dust models which is ``complete" subject to certain specified restrictions. 
This serves as a first step in the more general analysis, presented in an
accompanying paper (Carr
\& Coley 1999a), of spherically symmetric self-similar perfect fluid models with equation of state $p =\alpha \mu$. Despite the simplifications entailed in dropping
pressure, we will find that many of the features of the dust ($\alpha =0$) solutions carry over to 
the $\alpha \neq 0$ case, at least in the supersonic regime. In particular, many (though not all) of the types of solutions with pressure have direct analogues in
the dust case. Our aim is
therefore to use the exact analytic dust solutions to derive qualitative features of self-similar solutions which will also turn out to pertain when there is
pressure but which can then only be demonstrated numerically. It should, of course, be stressed that the introduction of
pressure is also associated with many new features - especially in the subsonic regime - which cannot be understood in this way. 

The plan of the paper is as follows: In Section 2 we will show how the dust equations can be regarded as a special case of the equations with pressure providing
one adopts a specific prescription in taking the limit $\alpha \rightarrow 0$.
In Section 3 we discuss the general family of self-similar solutions, showing that they can be conveniently categorized as
asymptotically Friedmann and what we term asymptotically quasi-static. In each case, we will focus on the important physical features of the solutions, such as
the presence of apparent horizons, event horizons and singularities. We will also clarify the connection between models with positive and negative
$z$. We will specify the sense in which our classification is ``complete"  in Section 4. 
    
\setcounter{equation}{0}
\section{Spherically Symmetric Similarity Solutions}

In the spherically symmetric situation one can introduce a time coordinate
$t$ such that surfaces of constant $t$ are orthogonal to fluid flow lines
and comoving coordinates ($r,\theta ,\phi$) which are constant along each
flow line. The metric can then be written in the form
\be
\label{lelement}
        ds^{2}=e^{2\nu}\,dt^{2}-e^{2\lambda}\,dr^{2}-R^{2}\,d\Omega^{2},\;\;\;
        d\Omega ^2 \equiv d\theta^{2}+\sin^{2}\theta \,d\phi^{2} 
\ee
where $\nu$, $\lambda$ and $R$ are functions of $r$ and $t$. The Einstein equations have a first integral
\be
\label{firstint}
   m(r)=\mbox{$\frac{1}{2}$} R\left[ 1+e^{-2\nu}
   \left(\frac{\partial R}{\partial t}
   \right)^{2} - e^{-2\lambda}\left(\frac{\partial R}
   {\partial r}\right)^{2}\right].
\ee
This can be interpreted as the mass within comoving radius $r$ at 
time $t$: 
\be
\label{massfunct}
   m(r)=4\pi\int_{0}^{r}\mu R^{2}\frac{\partial R}{\partial r'}\,dr'.
\ee
where $\mu(r,t)$ is the energy density and we choose
units in which $c=G=1$. This is constant if $p=0$.
Eqn (2.2) can be written as an equation for the energy per
unit mass of the shell with comoving coordinate $r$:
\be
\label{efunc}
   E\equiv\frac{1}{2}(\Gamma^{2}-1)
   =\frac{1}{2}U^{2}-\frac{m}{R},
    \;\;\;\;U\equiv e^{-\nu}\left(\frac{\partial R}{\partial t}\right),
   \;\;\;\;\Gamma\equiv e^{-\lambda}\left(\frac{\partial R}{\partial r}\right).
\ee
This can be interpreted as the sum of
the kinetic and potential energies per unit mass. $E$ and $\Gamma$ are conserved along fluid flow lines in the $p=0$ case. 

By a spherically symmetric similarity solution we shall mean one in 
which the spacetime admits a homothetic Killing vector \mbox{\boldmath$\xi$} that satisfies
\be
        \xi_{\mu ;\nu}+\xi_{\nu ;\mu} =2g_{\mu \nu}.
\ee 
This means that the solution is unchanged by a transformation of the form
$t\rightarrow at$, $r\rightarrow ar$ for any constant $a$. 
Solutions of this sort were first
investigated by Cahill \& Taub (1971), who showed that by a suitable
coordinate transformation they can be put into a form in which all
dimensionless quantities such as $\nu$, $\lambda$, $E$ and
\be
   S\equiv\frac{R}{r},\;\;\;\;M\equiv\frac{m}{R},\;\;\;\;
   P\equiv pR^{2},\;\;\;\;W\equiv\mu R^{2}
\ee
are functions only of the dimensionless variable $z\equiv r/t$. This means that the field equations reduce to a set of ordinary differential
equations in $z$. Another important quantity is the function
\be
\label{velocity}
        V(z)=e^{\lambda -\nu}z,
\ee
which represents the velocity of the surfaces of 
constant $z$ relative to the fluid. These surfaces have the equation $r=z t$ and therefore
represent a family of spheres moving through the fluid. The spheres contract relative to the fluid for $z<0$ and expand for $z>0$. This is to be distinguished 
from the velocity of the spheres of constant $R$ relative to the fluid:
\be
V_R = - \frac{U}{\Gamma} = - e^{\lambda -\nu} \left(\frac{\partial R/\partial t}{\partial R/\partial r}\right).
\ee 
This is positive if the fluid is collapsing and negative if it is expanding.
Special significance is attached to values of $z$ for which
$|V|=1$ and $|V_R|=1$. The first corresponds to a Cauchy horizon (either a black hole event horizon
or a cosmological particle horizon) and the second
to a black hole or cosmological apparent horizon. One can show that the existence of an apparent horizon is also equivalent to the condition $M=1/2$.

Although our main focus in this paper is (pressureless) dust solutions, it is elucidating to start by considering the equations for
a fluid with pressure. 
The only barotropic equation of state compatible
with the similarity ansatz is one of the form $p=\alpha \mu$ ($-1\leq\alpha\leq 1$). If one 
introduces a dimensionless function $x(z)$ defined by
\be
\label{xdef}
        x(z)\equiv (4\pi \mu r^{2})^{-\alpha/(1+\alpha)},
\ee
then the conservation equations $T^{\mu \nu}_{\;\;\;\; ;\nu}=0$ can be 
integrated to give
\be
\label{metrictt}
        e^{\nu}=\beta x z^{2\alpha/(1+\alpha)}
\ee
\be
\label{metricrr}
        e^{-\lambda}=\gamma x^{-1/\alpha} S^{2}
\ee
where $\beta$ and $\gamma$ are integration constants. The remaining field 
equations reduce to a set of ordinary differential equations in $x$ and 
$S$:
\be
   \ddot{S}+\dot{S}+\left(\frac{2}{1+\alpha}\frac{\dot{S}}{S}
   -\frac{1}{\alpha}\frac{\dot{x}}{x}\right)
   [S+(1+\alpha)\dot{S}]=0,
\ee
\be
   \left(\frac{2\alpha\gamma^{2}}{1+\alpha}\right)S^{4}
   +\frac{2}{\beta^{2}}\frac{\dot{S}}{S}\,x^{(2-2\alpha)/\alpha}
   z^{(2-2\alpha)/(1+\alpha)} - \gamma^{2}S^{4}\,\frac{\dot{x}}{x}
   \left(\frac{V^{2}}{\alpha}-1\right) = (1+\alpha)x^{(1-\alpha)/\alpha},
\ee
\be
   M=S^{2}x^{-(1+\alpha)/\alpha}\left[1+(1+\alpha)\frac{\dot{S}}{S}
   \right],
\ee
\be
   M=\frac{1}{2}+\frac{1}{2\beta^{2}}x^{-2}z^{2(1-\alpha)/(1+\alpha)}
   \dot{S}^{2}-\frac{1}{2}\gamma^{2}x^{-(2/\alpha)}S^{6}
   \left(1+\frac{\dot{S}}{S}\right)^{2},
\ee
where the velocity function is given by
\be
   V=(\beta\gamma)^{-1}x^{(1-\alpha)/\alpha}S^{-2}
   z^{(1-\alpha)/(1+\alpha)}
\ee
and an overdot denotes $zd/dz$. The other velocity function is
\be
V_R = \frac{V\dot{S}}{S+\dot{S}}\;\;
\ee
and, from eqns (2.4) and (2.15), the energy function is
\be
E = \frac{1}{2}\gamma^{2}x^{-(2/\alpha)}S^{6}
   \left(1+\frac{\dot{S}}{S}\right)^{2} - \frac{1}{2}\;\;,
\ee
which necessarily exceeds $-1/2$.

As discussed by Carr \& Yahil (1990), we can best
envisage how these equations generate solutions by working in the
$3$-dimensional $(x, S, \dot{S})$ space. At any point
in this space, for a fixed value of $\alpha$, eqns (2.14) and (2.15) give the value of z; eqn (2.13) then gives the value of $\dot{x}$ unless $|V|=\sqrt{\alpha}$ 
and
eqn (2.12) gives the value of $\ddot{S}$. Thus the equations generate a vector field $(\dot{x}, \dot{S}, \ddot{S})$ and this specifies an integral curve at each
point of the 3-dimensional space. Each curve is parametrized by $z$ and represents one particular similarity solution.
This shows that, for a given equation of state parameter $\alpha$, there is a
$2$-parameter family of spherically symmetric similarity solutions. In general there would be a sonic point, with possible associated discontinuities,
at $|V|=\sqrt{\alpha}$. This corresponds to crossing a 2-dimensional surface in the solution space but we need not discuss this complication here. 

We now focus on the dust solutions. Although eqns
(2.10) to (2.16) break down when $\alpha=0$, in that some of the terms appearing there disappear or diverge, we will show that they are still formally
applicable providing the function $x$ defined by eqn (2.9) is set to $1$ whenever it does not appear with the exponent $1/\alpha$. 
Otherwise one must make the substitution
\be
x^{1/\alpha} \rightarrow (4\pi \mu r^2)^{-1},\;\;\;\frac{1}{\alpha}\frac{\dot{x}}{x} \rightarrow -\frac{dln(\mu r^2)}{dlnz},
\ee
as suggested by eqn (2.9). In fact, the relevant equations are most simply obtained by noting that both
the energy and mass within comoving radius $r$ are conserved, so that $E$ and $m/r=MS$ are constant. If
we put $m/r=\kappa$, then eqn (2.3) implies
\be
\label{density}
 4\pi\mu R^{2}\frac{\partial R}{\partial r}=\frac{dm}{dr}=\kappa
\ee
and this can be combined with eqns (2.9) and 
(2.11) to give
\be
\label{ssgrr}
   e^{\lambda}=\gamma^{-1}(4\pi\mu r^{2}S^{2})^{-1} =
   \kappa^{-1}\gamma^{-1}\frac{\partial R}{\partial r}.
\ee
On the other hand, eqn (2.4) implies
$e^{\lambda}=\Gamma^{-1}(\partial R/\partial r)$ where $\Gamma = \pm\sqrt{1+2E}$ and so the 
constant $\kappa$
is just $\Gamma/\gamma$. The mass function is therefore
\be
\label{ssmass}
   M=\frac{\Gamma}{\gamma S}=\frac{\sqrt{1+2E}}{\gamma S},
\ee
where we have taken the positive square root for $\Gamma$ to ensure that the mass if positive. (We discuss the negative mass case later.) Putting $x=1$ in eqn
(2.10) also gives $e^{\nu}=\beta$, so eqn (2.4) can be rewritten as 
\be
\label{ssenergy}
   E=\frac{1}{2\beta^{2}}z^{4}\left[\frac{dS}{dz}\right]^2
   -\frac{\sqrt{1+2E}}{\gamma S}.
\ee
Once this equation has been integrated to give $S(z)$, all the other functions can be obtained, so one has solved the problem completely. 

We now show that eqns (2.10) to (2.16) are all formally satisfied if one uses the prescription given by eqn (2.19). Eqn (2.14) can be written as
\be
MS = m/R = S^2(4\pi\mu r^2)(S+\dot{S}) = 4\pi\mu R^2\frac{\partial R}{\partial r}
\ee
and this is just equivalent to eqn (2.20). Since eqn (2.21) implies
\be
(4\pi \mu r^2)^{-1} = \gamma S^2(S+\dot{S})/\Gamma,
\ee
we also have
\be
\frac{dln(\mu r^2)}{dlnz} = - \frac{2\dot{S}}{S} - \frac{\dot{S}+\ddot{S}}{S+\dot{S}}\;\;.
\ee
Eqn (2.12) is then automatically satisfied and from eqn (2.23) formally corresponds to $\dot{E}=0$. 
Finally we can substitute for the $x$ terms in eqn (2.13), using eqns (2.19) and (2.25), to obtain
\be
\ddot{S} + \dot{S} = - \frac{\Gamma}{S^2z^2}
\ee
and this can be integrated to give eqn (2.23). 

It is now convenient to 
scale the $r$ and $t$ coordinates so that $\beta=\gamma=1$. Eqn (2.23) then implies 
\be
\label{ssdSdz}
   \frac{dS}{dz} = \pm \frac{\sqrt{2E+2\Gamma/S}}{z^{2}}
\ee
and this can be integrated to give 
\be
\label{ssdust}
   D\mp \frac{1}{z}=\left\{ \begin{array}{ll}

   \frac{\sqrt{ES^{2}+\Gamma S}}{\sqrt{2}E}-
\frac{2\Gamma}{(2E)^{3/2}} 
   \sinh^{-1}\sqrt{\frac{ES}{\Gamma}} & (E>0) \\ & \\

   \frac{\sqrt{2}}{3}S^{3/2} & (E=0) \\ & \\

   \frac{2\Gamma}{(-2E)^{3/2}}
   \sin^{-1}\sqrt{\frac{-ES}{\Gamma}} \pm \frac{\sqrt{ES^{2}+\Gamma 
S}}{\sqrt{2}E} & (-1/2<E<0)

   \end{array} \right.
\ee
where D is an integration constant and $\sin^{-1}$ is taken to lie between
$0$ and $\pi$. In the first two cases, the upper and lower signs on the left apply for $dS/dz$
positive and negative, respectively, this sign being constant for any particular solution. In the third case, the sign on the left is fixed for a 
particular solution, even though $dS/dz$ may switch
sign, but the sign of the last term is
plus if $\sin^{-1}$ lies between $0$ and $\pi/2$ and minus if it lies between $\pi/2$ and $\pi$. If we took the negative square
root in eqn (2.22), corresponding to $M$ and
$\Gamma$ being negative, there would be another solution for $E>0$ given by
\be
D\mp \frac{1}{z}= \frac{\sqrt{ES^{2}-|\Gamma| S}}{\sqrt{2}E}+
\frac{2|\Gamma|}{(2E)^{3/2}} 
   \cosh^{-1}\sqrt{\frac{ES}{|\Gamma|}} \;\;\; (E>0). 
\ee
This solution is unphysical, since the mass is negative, but it is of interest for comparison with the solutions with pressure. 

Eqns (2.16), (2.17), (2.19), (2.25) and (2.28) give the
velocity functions as
\be
\label{ssvel}
   V=\frac{Sz\pm\sqrt{2E+2\Gamma /S}}{\Gamma}
\ee
and
\be
\label{ssvelR}
V_R= \pm\frac{\sqrt{2E+2\Gamma /S}}{\Gamma},
\ee
while eqns (2.16) and (2.31) imply that the density is given by
\be
4\pi \mu t^2 = \frac{1}{zS^2V} = \frac{\Gamma}{zS^2(Sz\pm\sqrt{2E+2\Gamma /S})},
\ee
where the upper and lower signs again correspond to $dS/dz$ being positive and negative, respectively. 
Note that V and $\mu$ is negative (corresponding to tachyonic
models) for the solution given by eqn (2.30) and $\mu$ is also negative. 
In all cases the metric can be written as
\be
ds^2 = dt^2 - \frac{(S+\dot{S})^2}{1+2E}dr^2 - r^2 S^2d\Omega^2,
\ee
which is the standard Tolman-Bondi form with constant energy function $E(r)$.

\section{Classification of Solutions}

Eqn (2.29) implies that there is a
2-parameter family of similarity solutions (as in the general 
$\alpha$ case). In this section we will provide a complete description of these solutions. We will start by considering the simplest
one: the flat Friedmann solution with
$D=E=0$. We will then consider the one-parameter family of solutions with $E\neq 0, D=0$. Finally we will consider the full two-parameter family of solutions
with  $E\neq 0, D\neq 0$. In each case, we will show the form of the physically interesting quantities $S$, $V$
and $\mu t^2$ as functions of $z$. In obtaining the full family of solutions, it is crucial that we allow $z$ to be either positive or negative. 
Our analysis will also cover the (presumably unphysical) solutions with negative mass because they relate to some of the solutions with pressure.

\subsection{$E=D=0$ solution}

In this case eqns (2.21), (2.29), (2.31) and (2.33) give
\be
S=(\sqrt{2}z/3)^{-2/3}=M^{-1},\;\;\;V=(z/6)^{1/3},\;\;\;\mu = (6\pi t^2)^{-1}.
\ee 	
This corresponds to the standard dust Friedmann model with zero curvature constant. The metric can be put
in the usual form by making the substitution $\hat{r}=(9r/2)^{1/3}$, which gives
\be
ds^2= dt^2 - t^{4/3}[d\hat{r}^2 + \hat{r}^2 d\Omega^2].
\ee
Note that the curvature constant must be zero because otherwise there would be an intrinsic scale, which would contradict the similarity assumption.
   
\subsection{$D=0$ solutions}

Solutions with $D=0$ are asymptotically Friedmann as 
$|z|\rightarrow \infty$ and are specified entirely by the energy 
parameter $E$ and were studied by Carr
\& Hawking (1974) and Carr \& Yahil (1990). The form of $S(z)$ in these solutions is shown in Figure (1a), the arrows always corresponding
to the direction of increasing time. The solutions with $z>0$
correspond to initially expanding Big Bang models: they start from an initial Big Bang singularity ($S=0$) at $t=0$ ($z=\infty$) and then
either expand indefinitely ($S\rightarrow \infty$) as $t \rightarrow \infty$ ($z \rightarrow 0$) for $E\ge 0$ or recollapse to a black hole singularity ($S=0$) at
\be
z_S=\frac{(-2E)^{3/2}}{2\pi\sqrt{1+2E}}
\ee
for $E<0$. Note
that $z_S$ corresponds to the physical origin since $R=rS=0$ there. The form of $V(z)$ in the $z>0$ solutions is shown in Figure (1b). In the first case, $V$
decreases monotonically from $\infty$ to 0. In the second case, it reaches a minimum before rising to $\infty$ at $ z_S$. One can show that the values of $z$ and $V$
at the minimum both decrease as $E$ increases; the minimum will exceed 1 (in which case the whole Universe is inside the black hole) if $E$ is less than some critical
negative value $E_*$ and it will be less than 1 (in which case there is a black hole event horizon and a cosmological particle horizon) if $E$ exceeds $E_*$. The
solutions with $z<0$ are the time-reverse of the $z>0$ ones and the sign of $V$ is also reversed: as $t$ increases from $-\infty$ to 0 (i.e.
as
$z$ decreases from 0 to
$-\infty$), the
$E\ge 0$ models collapse from an infinitely dispersed state ($S=\infty$) to a Big Crunch singularity ($S=0$); the
$E<0$ models also collapse to a Big Crunch singularity but they emerge from a white hole and are never infinitely dispersed.

Both $S$ and $V$ have the same $z$-dependence as in the $E=0$ Friedmann solution as $|z|\rightarrow\infty$:
\be
S \approx [9\sqrt{1+2E}/2]^{1/3}|z|^{-2/3},\;\;\;V \approx [6(1+2E)]^{-1/3}z^{1/3}. 		
\ee
However, the $E\neq 0$ solutions deviate from the $E=0$ solution at small values of $|z|$. The $E<0$ solutions never reach $z=0$ at all, while the $E>0$ ones have 
\be
S \approx (2E)^{1/2}|z|^{-1} , \;\;\;
V \approx -(1+2E)^{1/2} E^{-1}z\,\ln[(2E)^{3/2}(1+2E)^{1/2}|z|]\label{eq:  }
\ee
as $|z| \rightarrow 0$. The first relation implies that the circumference function $R(r,t)=Sr$ is non-zero in limit $r\rightarrow 0$ unless $E=0$ since
\be
R(0,t) = \sqrt{2E}\, t.\label{eq:  }
\ee
This means that the ``coordinate'' origin ($r=0$) is an 
expanding 2-sphere. [This feature is specific to the dust case and does not arise if there is pressure.] 
This has a natural physical interpretation since the forms of
$S$ and
$V$ are similar to those in the Kantowski-Sachs (1966) solution,  in which
all the matter is localized on a shell [cf. eqns (3.19) and (3.23) in Carr \& Coley (1999a)]. However, we note that there is no exact self-similar 
Kantowski-Sachs solution in the
dust case. To obtain a complete solution, one must therefore match the self-similar solution onto a (non-self-similar) part inside $R(t,0)$. In the $E<0$ case,
we  have seen that the physical origin is the black hole
singularity $z_S$, so only for $E=0$ can one identify $z=0$ with the physical origin.

The form of the density function $\mu t^2$ can be derived from eqn (2.33) and is shown in Figure (1c). For a given fluid element, this specifies the density as
a function of time $\mu(t)$. For a given time, it also
specifies the density profile
$\mu(r)$ and this illustrates that a  non-zero value of $E$ necessarily introduces an 
inhomogeneity into the model. Solutions with $E>0$ are everywhere underdense relative to the Friedmann model, with eqns (2.33) and (3.5) implying that 
$\mu t^2$ goes to 0 as $(\ln|z|)^{-1}$ 
as $z\rightarrow 0$. (This  suggests that the interior non-self-similar region should be a vacuum.) Solutions with
$E<0$  are everywhere overdense relative to Friedmann, with $\mu$ diverging at the singularity. Note that eqns (2.33) and (3.4) imply that $\mu t^2$ is
independent of $E$ to 1st order as $|z| \rightarrow \infty$.

The form of the mass function $M(z)$ 
in the $D=0$ solutions is not shown explicitly but can be immediately deduced from the expression for $S$ since eqn (2.21) gives $M=\Gamma/S$. 
In the $E\geq 0$ case, there is always a single point where $M=1/2$ and this corresponds to the cosmological apparent horizon. In the $E<0$ case, eqn (2.29) implies 
that $S$ has a maximum of $\Gamma/|E|$ and so eqn (2.21) shows that 
$M$ has a minimum 
of $|E|$. Since this is less than 1/2, there are always two points where
$M=1/2$, one corresponding to the 
black hole apparent horizon and the other to the cosmological apparent horizon.
Note that a black hole's apparent horizon always lies within or coincides with its event horizon (Hawking \& Ellis 1973),
which is why the first can exist without the second. Eqns (2.21) and (3.3) imply that the mass associated with this singularity is
\be
m_S=(MSz)_S\;t = (-2E)^{3/2}t /(2\pi).
\ee
It therefore starts off zero when the singularity first forms at $t=0$ but then grows as $t$. The mass of the black hole is given by a similar
formula but with $z$ having the value appropriate for the event horizon or apparent horizon. Since the former may not exist,
it is more appropriate to use the latter. 

Finally, we consider the $E>0$ negative-mass solutions given by eqn (2.30). Their form is indicated by the dotted curves in Figures (1). $S$, $V$ and $\mu$ have the
same form as in the positive mass solutions for small values of $|z|$ except that V and $\mu$ reverse their signs.  However, the solutions are very different at
large values of $|z|$ since eqn (2.30) shows that $S$ must always exceed $|\Gamma|/E$. Indeed it tends to this value asymptotically, so we have
\be
S\approx \sqrt{1+2E}/E,\;\;\;V \approx -z/E,\;\;\;\mu r^2 \approx -E^3/\sqrt{1+2E}
\ee
as $|z| \rightarrow \infty$. The form of this solution is closely related to that of the $\alpha <<1$ static solution [cf. eqn (3.29) of Carr \& Coley (1999a)],
although there is no static solution in the $\alpha=0$ case itself. 

\subsection{$E=0$ solutions}

We now put $E=0$ and consider the effect of introducing a non-zero value for the constant $D$. [In this case, eqn (2.28) does not permit $\Gamma <0$, so there are 
no negative-mass solutions.] The form of $S(z)$ for the $D>0$ solutions is shown in Figure (2a). There are two types of solutions in this case, one expanding and the
other collapsing. For the expanding solutions (solid lines), $S=0$ at $z=-1/D$ and so the Big Bang occurs before $t=0$ (i.e. it is ``advanced''). As $t$ increases to
0 (i.e. as $z$ decreases to $-\infty$), $S$ tends to the finite value
\be
S_\infty(D)=(3D/\sqrt{2})^{2/3}.
\ee
As $t$ further increases from 0 to $+\infty$ (i.e. as $z$ jumps to 
$+\infty$ and then decreases to 0), $S$ increases monotonically to $\infty$. For the contracting solutions (broken lines), $S$ starts infinite at $t=-\infty$ 
($z=0$) and then decreases to $S_\infty(D)$ as $t$ increases to 0 ($z \rightarrow -\infty$). As $t$ further increases (i.e. as $z$ jumps to $+\infty$ and then
decreases), $S$ continues to decrease until it reaches 0 at the Big Crunch singularity at $z=1/D$. Both types of solutions are characterized by the fact that they
have just one singularity and span both positive and negative values of $z$. Note that for each value of $S_\infty$ the two asymptotic solutions just correspond to
the plus and minus signs in eqn (2.28).

The form of $V(z)$ in the $D>0$ solutions is 
shown in Figure (2b). For the expanding solutions (solid lines), it starts off at $-\infty$ 
at the Big Bang ($z=-1/D$), reaches a negative maximum and then, from eqn (2.31), tends to 
\be
V = S_\infty(D) z = (3D/\sqrt{2})^{2/3} z
\ee
as $z\rightarrow -\infty$. When z jumps $+\infty$, $V$ becomes positive but eqn (3.10) still applies. As
$z$ decreases from $+\infty$ to 0, $V$ decreases monotonically to 0. For the contracting models (broken lines), $V$ starts from zero at
$z=0$ and monotonically decreases as $z$ goes to $-\infty$, being again given by eqn (3.10) asymptotically. When $z$ jumps to $+\infty$, $V$ jumps to $+\infty$
and then decreases to a minimum before rising to infinity  at the Big Crunch singularity. Note that the maximum value of $V$
for the expanding  solutions will exceed $-1$ and the minimum value for the contracting ones will be less than $+1$ (i.e. the minimum value of $|V|$ is
less than $1$) if
$D$ exceeds some critical value $D_+$. A simple calculation (see later) shows that the values of $|z|$ and $|V|$ at the stationary point are given by 
\be
|z|_{min} = \left(\frac{2+\sqrt{3}}{3D}\right),\;\;\;|V|_{min} = \left(\frac{26+15\sqrt{3}}{3D}\right)^{1/3},
\ee
so the stationary point moves towards the origin as $D$ increases and $D_+ = 26/3+5\sqrt{3} \approx 17$. For $D>D_+$, the condition
$|V|=1$ will be satisfied at {\it three} values of
$z$. As illustrated by  Figure 14 of Ori \& Piran (1990), this means
that the contracting solutions will form a black hole in which the central singularity is naked for a while.  

The crucial feature of these solutions is that, while the form of $V(z)$ is like that in the $D=0$ case for small $|z|$, $V$ scales as $z$ rather
than
$z^{1/3}$ [cf. eqn (3.4)] for large $|z|$. This is because any solution with finite $S$ at infinity must be ``nearly'' static in the sense that
$dS/dz$ tends to zero. However, the  solutions are not asymptotic to an {\it exact} static solution (indeed this does not exist in the $\alpha=0$ case) because eqn
(2.32) implies that $V_R$ tends to a non-zero value:
\be
V_R^{\infty} = \pm \left(\frac{4}{3D}\right)^{1/3}.
\ee
We therefore term these solutions asymptotically 
``quasi-static''. If $V_R^{\infty}$ is positive, the fluid is collapsing at infinity; if $V_R^{\infty}$ is negative, it is expanding. Note that eqn (2.28) implies
that both
$dS/dz$ and 
$zdS/dz$ tend to zero at large $|z|$ but
$z^2dS/dz$ [which directly relates to $V_R^{\infty}$ from eqns (2.28) and (2.32)] tends to a non-zero value except in the limit $D\rightarrow \infty$.  

The form of the density function $\mu t^2$ in the $D>0$ solutions is also interesting and is illustrated in Figure (3c). From eqn (2.33) the density parameter is
given by
\be
\Omega \equiv 6\pi\mu t^2 = \frac{1}{(1\mp 3Dz)(1\mp Dz)}
\ee
where the upper and lower signs apply for positive and negative values of $dS/dz$, respectively. [The inverse of the factor $6\pi t^2$ 
corresponds to the density in a
flat Friedmann dust  universe, as indicated by eqn (3.1).] For a given fluid element, this describes how the density evolves as a function of time and it has the
expected form. However, at a given time it also prescribes the density profile and one sees immediately that a non-zero value of $D$ (like a non-zero value of $E$)
introduces an inhomogeneity.  This inhomogeneity has a particularly interesting form. In the $z<0$ regime, the profile for the collapsing solutions is homogeneous
for $|z|<<1/D$ but has $\mu \sim r^{-2}$ for $|z|>>1/D$ (i.e. it resembles an isothermal sphere with a uniform core). In the $z>0$
regime, the collapsing solutions again have 
$\mu \sim r^{-2}$ for $|z|>>1/D$ but the density diverges at $z=1/D$  (i.e. one has a density singularity at the centre of an isothermal sphere). 
For the expanding solutions, the signs of $z$ are reversed. 
These features are illustrated in Figure (2c) and have an obvious physical interpretation.
 
Although the asymptotically quasi-static solutions have an natural cosmological interpretation when $z$
is allowed to span both positive and negative values, we see that the $z>0$ and $z<0$ solutions also have a non-cosmological 
interpretation when considered separately: they just represent collapsing and expanding self-similar models which evolve from an initially isothermal 
distribution. It
is interesting that the isothermal model (which is usually associated with a static solution) features
prominently in both regimes, despite the fact that there is no exact static solution in the dust case. Note that the behaviour at $z=0$
is different from the $D=0$ case, in that $\Omega$ tends to $1$ rather than $0$ for $E>0$. 

The mass function is $M=S^{-1}$ in this case and therefore decreases or increases monotonically. There is just one value of $z$ at which $M=1/2$ 
(corresponding to a black hole or cosmological apparent horizon) but this
may be in either the positive or negative $z$ region. Eqn (3.9) implies that the asymptotic value of $M$ as $|z|\rightarrow \infty$ is less than 
$1/2$  for $D>4/3$. In this case, the collapsing solutions have their apparent horizon in $z>0$, whereas the expanding ones have 
it in $z<0$. The mass of
the (possibly naked) singularity in these solutions is
\be
m_S=(MSz)_S\;t = t /D
\ee
from eqn (2.21). As in the asymptotically Friedmann case, it starts off zero at $t=0$ but then grows as $t$. In the limit $D\rightarrow \infty$, one gets a naked
singularity of zero mass at the origin (cf. the  static solution).

Finally, we consider the $D<0$ solutions. The form of $S(z)$ in this case is shown by the dotted curves in Figure (2a). Such solutions are confined to $|z|<-1/D$, 
with $S$ either decreasing monotonically for $z<0$ (i.e. as $t$ increases from $-\infty$) or increasing monotonically for $z>0$ (i.e. as $t$ increases to $+\infty$).
However, these solutions break down when $S$ is too small. This is because eqn (2.31) implies that $|V|$
increases to some maximum value and then falls to zero at $|z|=-1/(3D)$; this is indicated by the dotted curve in Figure (2b). From eqn (2.33) this means that the
density diverges there, as shown by the dotted curve in Figure (2c). This divergence is associated with the formation of a shell-crossing singularity since the
model resembles the Kantowski-Sachs solution at this point. For
$-1/D>|z|>-1/(3D)$, the density and velocity functions become negative but this is presumably unphysical. 

\subsection{$D\neq 0, E\neq 0$ solutions}

The forms of $S(z)$ for the $(D>0, E\neq0)$ solutions are indicated in Figure (3a). The figure assumes that $D$ is 
fixed but allows $E$ to vary. The $(D>0, E>0)$ solutions are qualitatively similar to the $(D> 0, E=0)$ ones in that they are 
monotonically expanding or collapsing and span both positive and negative $z$, 
as illustrated by the upper solid and broken curves, respectively. They are also asymptotically quasi-static, in the sense that
$S$ tends to a finite value as $|z|\rightarrow \infty$, even though $V_R$ is non-zero there from eqn (2.32). The form of the
solutions near $z=0$ is still given by eqns (3.5) for $E>0$, so the
behaviour is like that in the $(E>0, D=0)$ case here. In particular, $z=0$ no longer corresponds to the physical
origin, so one again has to attach the solution to a non-self-similar central region.

The $(D> 0, E<0)$ solutions are qualitatively different
from the $(D> 0, E=0$) ones in that the models no longer collapse from or expand to infinity. This is clear
from eqn (2.28), which implies that $S$ has a maximum value of $\Gamma/|E|$, so all the solutions start off 
expanding and then recollapse. Note that there
is no exact static solution since that would be incompatible with eqn (2.13), the term on the right-hand-side being
non-zero for $\alpha =0$. Figure (3a) shows that there are two
types of $(D> 0, E<0)$ solutions. One type (illustrated by the lower solid curves) expands from the Big Bang singularity at
$z=-1/D$ and then recollapses to a black hole singularity at 
\be
z_S=\left[\frac{2\pi\sqrt{1+2E}}{(-2E)^{3/2}} -D \right]^{-1}.
\ee
This reduces to the value given by eqn (3.3) if $D=0$. The other type (illustrated by the lower broken
curves) expands from a
white hole singularity at $-z_S$ and then recollapses to a Big Crunch singularity at $z=1/D$. 

Eqn (3.15) implies that $z_S=1/D$, so that the solution is symmetric in $z$, if $D$ has the value
\be
D_{sym} \equiv \frac{\pi\sqrt{1+2E}}{(-2E)^{3/2}}\;.
\ee
One can invert this condition to obtain the associated value of $E$ in terms of $D$:
\be
E_{sym} \equiv - \frac{4\pi}{\sqrt{3}D} \sinh \left[ \frac{1}{3}\sinh^{-1}\left(\frac{3\sqrt{3}D}{8\pi}\right)\right]
\ee
and this specifies a $1$-parameter family of solutions with $V_R^{\infty} =0$. If one
considers the limit of the symmetric solution as $D\rightarrow0$, one finds $E_{sym}\rightarrow -1/2$ and $z_S\rightarrow\infty$. On the other hand,
if one
considers the limit as $D\rightarrow\infty$, one finds $E_{sym}\rightarrow 0$ and $z_S\rightarrow0$, so that both singularities 
go the
origin. This is the  closest one
can get to a static solution in the dust case. Note also that $z_S\rightarrow 0$
in the limit $E\rightarrow0^-$ whatever the value of $D$; the sudden transition as one goes from $E=0^-$ to $E=0^+$ is illustrated in Figure (3a).

This value of $E$ given by eqn (3.17) has a special physical significance in that it prescribes the {\it minimum} value of $E$ allowed for
given $D$, as indicated by the lower boundary in Figure (4). The proof of
this is as follows. If one takes the limit of eqns (2.14) and (2.15) with $\alpha=0$ as $z\rightarrow \infty$, using eqns (2.19) and (2.32) and the fact that
$\dot{S}\rightarrow 0$ from eqn (2.28), one obtains
\be
z^2\dot{S}^2 = (1+2E)(V_R^\infty)^2 = (4\pi\mu r^2)_{\infty}^2 S_{\infty}^6 + (8\pi \mu r^2)_{\infty}S_{\infty}^2 - 1
\ee
where one can regard $(\mu r^2)_{\infty}$ and $S_{\infty}$ as independent asymptotic parameters. If one now fixes $(\mu r^2)_{\infty}$ and assumes
that it is positive, then the
right-hand-side of eqn (3.18) decreases monotonically with decreasing $S_{\infty}$. One therefore gets a {\it real} solution for $V_R^\infty$ only if $S_{\infty}$
exceeds a certain value and - by monotonicity - this must be the
value associated with  the symmetric solution. Thus a real solution (with positive density) exists only for $E>E_{sym}$ or, equivalently, $D<D_{sym}$. This means 
that $z_S$ is always positive and less than
$1/D$ and that the maximum of
$S$ will always occur at the opposite sign of
$z$ as the
$|z|=1/D$ singularity. These features are illustrated by the curves in Figure (3a). 

The form of $V(z)$ in the these solutions is shown in Figure (3b). For $(D> 0, E>0)$ it is similar to that in the $(D> 0, E=0)$ case. As $z$ decreases
from $0$ to $-\infty$, $V$
decreases monotonically from $0$ to $-\infty$; it then jumps to $z=+\infty$ and, as $z$ continues to decrease, it falls to a minimum and
rises to infinity at the Big Crunch singularity at $z=1/D$. Also as in the
$(D> 0, E=0)$ case, this minimum will fall below 1, corresponding to a naked
singularity, providing $D$ exceeds some value $D_+(E)$. We derive an implicit expression for $D_+(E)$ later but, for the present, 
note that it increases with increasing
$E$ and reduces to the value 
$D_+$ which arose in Section 3.3 when $E=0$. The condition for a naked singularity can
also be expressed as the requirement that $E$ exceed some critical value $E_+(D)$.

For $(D> 0, E<0)$ the form of
$V$ is similar to that in the
$(D= 0, E<0)$ case. As $z$ decreases from $-1/D$, $V$ rises from
$-\infty$  until some maximum and then falls to $-\infty$ quasi-statically as $z\rightarrow -\infty$. It then jumps to $z=+\infty$ and falls to a minimum before
rising to
$\infty$ at $z_S$. As in the $(D=0, E>0)$ case, there will be a black hole event horizon if the minimum value of $V$ is less than $1$; this requires that $E$ exceed
some value
$E_*(D)$, which must reduce to the value $E_*$ given in Section 3.2 when $D=0$. We derive an implicit expression for $E_*(D)$ below. 
It should be stressed that the value  $E_*(D)$ is
associated with the minimum of
$|V|$ near the singularity at $|z|=z_S$ and is different from the value $E_+(D)$ associated with the minimum near 
$|z|=1/D$. The relationship between the values $E_+(D)$ and $E_*(D)$ is discussed below. 

In order to understand the form of the curves in Figures (3a) and (3b) more precisely, it is useful to specify their asymptotic behaviour. As
$|z|\rightarrow\infty$, eqns (2.29) and (2.31) with $E>0$ imply that $S(z)$ and $V(z)$ have the following asymptotic forms:
\be
S_{\infty} \approx  D\sqrt{2E}, \;\;\; V_{\infty} \approx \left(\frac{2E}{1+2E}\right)^{1/2}D z
\ee
for $D>>[(1+2E)/E^3]^{1/2}$ and
\be
S_{\infty} \approx \left(\frac{1+2E}{4}\right)^{1/6}(3D)^{2/3},\;\;\; V_{\infty} \approx \frac{(3D)^{2/3}z}{2^{1/3}(1+2E)^{1/3}}
\ee
for $D<<[(1+2E)/E^3]^{1/2}$. The transition value for $D$ between these two regimes is just an extrapolation of the expression for 
$D_{sym}$ given by eqn (3.16) into the $E>0$ regime; it scales as $E^{-1}$ for $E>>1$ and $E^{-3/2}$ for $E<<1$. Note that eqn (3.20) agrees with
eqns (3.9) and (3.10) in the limit $E=0$. For
$E<0$, eqn (3.20) still applies if $D<<D_{sym}$ but one has
\be
S_{\infty} \approx  \sqrt{1+2E}/|E|, \;\;\; V_{\infty} \approx z/|E|
\ee
for $D \approx D_{sym}$ (i.e. $S_{\infty}$ tends to the value associated with the symmetric solution). 

These equations prescribe the asymptotic forms for $S_{\infty}$ and $V_{\infty}$ in the different $(E,D)$ regimes of Figure 4. For fixed $D$, 
eqns (3.19) to (3.21) show that
$S_{\infty}$ always increases with $E$ but is roughly constant for 
$|E|<<1$. The behaviour of $V_{\infty}$ is more complicated: for $D<<1$, it first decreases 
with increasing $E$ and then flattens off; for $D>>1$, it first increases with increasing $E$ before flattening off. These features are indicated
in the  Figures (3a) and (3b). Note that all these solutions are
quasi-static and not exactly static asymptotically since eqn (2.32) gives
\be
V_R^{\infty} \approx \pm \frac{4^{1/3}}{(3D)^{1/3}(1+2E)^{1/3}},\;\;\;V_R^{\infty} = \pm \left(\frac{2|E|}{1+2E}\right)^{1/2} 
\ee
for $D >> [(1+2E)/E^3]^{1/2}$ and $D << [(1+2E)/|E|^3]^{1/2}$. The sign is positive for collapsing solutions and negative for expanding ones.
Note that the first expression agrees with eqn (3.12) in the limit $E=0$.

We now derive implicit expressions for the functions $E_*(D)$ and $E_+(D)$, which are related to the existence of event horizons or naked singularities.
Differentiating eqn (2.31) shows that when $dV/dz=0$ one always has
\be
V=1/(S^2z)
\ee
and eqn (2.31) then gives
\be
V\Gamma-1/(VS) = \pm \sqrt{2E +2\Gamma/S}\;\;,
\ee
where the positive and minus signs corresponds to the sign of $dS/dz$. If one also requires $|V|=1$ at the stationary point, eqn (3.24) gives two roots 
\be
S=2\Gamma \pm \sqrt{4\Gamma^2-1},\;\;\;|z|=8\Gamma^2 -1 \mp 4\Gamma\sqrt{4\Gamma^2-1},
\ee
where the plus and minus signs are distinct from the ones appearing in eqn (3.24). These roots can be real providing $\Gamma >1/2$, corresponding to $E>-3/8$.
However, one needs to check whether both of these solutions satisfy condition (3.24). 

Inserting the solutions (3.25) into eqn (2.29) gives an expression for
$D$ in terms of $E$. This expression is complicated in general but it simplifies in certain regimes. 
For $E>>1$, which implies $\Gamma \approx \sqrt{2E}>>1$, one
obtains two possible solutions:
\be
S\approx 4\Gamma,\;\;\; |z|\approx 1/(16\Gamma^2),\;\;\;D\approx 32E
\ee
and
\be
S\approx 1/(4\Gamma),\;\;\; |z|\approx 16\Gamma^2,\;\;\;D\approx \frac{1}{E}\left(\frac{13}{32} - \ln \sqrt{2}\right) \approx 0.06E^{-1}.
\ee
The first has $\Gamma >1/S$ and therefore requires $dS/dz>0$ from eqn (3.24), which leads to a consistent solution. However,
the second has $\Gamma <1/S$ and requires $dS/dz<0$, which does not. In the limit $E\rightarrow 0$, which implies $\Gamma \approx 1$,
one obtains 
\be
S\approx (2\pm \sqrt{3})(1 \pm 2E/\sqrt{3}),\;\;\;|z| \approx (7 \mp 4\sqrt{3})(1 \mp 4E/\sqrt{3}). 
\ee
The upper sign gives $\Gamma >1/S$ and therefore requires $dS/dz>0$, which leads to
a consistent solution as $E$ tends to
$0$ from either above or below. Eqn (2.29) then gives
\be
D \approx \left(\frac{26+15\sqrt{3}}{3}\right) +  E \left(\frac{109 + 63\sqrt{3}}{6}\right) \approx 17 + 36 E.
\ee
Note that the constant part of this expression gives the same limiting value of
$D$ as implied by eqn (3.11). The lower sign in eqn (3.28) gives 
$\Gamma <1/S$ and  requires $dS/dz<0$, which does not lead to a consistent solution. Finally we note that eqns (3.25) and (2.29) 
lead to a unique value of $E$ and $D$ for which the symmetric solution has $V_{min}=1$; this necessarily corresponds to a point on the lower
boundary in Figure 4. 

These limiting behaviours allow one to infer the rough form of the functions $E_*(D)$ and $E_+(D)$, as
indicated in Figure (4). Here we have used the fact that $dE_+/dD$ is positive as $E \rightarrow 0$, as follows from eqn (3.29). The form of the functions in the
$E<0$ regime can be inferred from the fact that $E_*(D)$ must reach the value 
$E_*$ mentioned in Section 3.2 when $D=0$ (although this value has not been calculated explicitly). Also $E_+(D)$ and $E_+(D)$ must reach the line $E=E_{sym}(D)$ at
the same value of
$E$ and this must clearly exceed $-3/8$ from eqn (3.25). We note that, for sufficiently large values of $D$, there may be {\it both} an
event horizon and a naked singularity.

The form of $\mu t^2$ in these solutions is shown in Figure (3c), although this gives only some of the solutions shown in Figures (3a) and (3b). It can be understood
as a composite of the curves shown in Figure (1c) for
$E<0$ and Figure (2c) for $E>0$. From eqns (2.33) and (3.19) to (3.21), the asymptotic form of the density profile density is given by
\be
4\pi \mu r^2 \approx \left[\frac{1+2E}{D^3(2E)^2},\;\;\;\frac{2}{9D^2}\right]
\ee
for  $D >> [(1+2E)/E^3]^{1/2}$ and $D << [(1+2E)/|E|^3]^{1/2}$ [cf. eqns (3.8) and (3.13).] The $E>0$ solutions are everywhere
underdense relative to the $E=0$ solutions, going to $0$ as
$(\ln|z|)^{-1}$ at the origin, whereas the
$E<0$ solutions are everywhere overdense and have a second density singularity. There is a uniform core region only in the $E=0$ case, although CC show that this
also applies for $E<0$ if there is pressure. 

The form of $M(z) = \Gamma/S(z)$ can be deduced immediately from Figure (3a). In the $E>0$ case it rises or falls monotonically, as in the $D=0$ case, so there is
just one value of $z$ for which $M=1/2$. As $|z|\rightarrow\infty$, $M$ tends to a limiting value
\be
M_\infty = \frac{\sqrt{1+2E}}{S_\infty(D,E)}
\ee
and the apparent horizon will be in $z>0$ or $z<0$ according to whether this is greater or less than 1/2. In the $E<0$ case, 
$M$ will have a minimum where $S$
has a maximum. This occurs at 
\be
|z|= \left[\frac{\pi\sqrt{1+2E}}{(-2E)^{3/2}}-D\right]^{-1}
\ee
and since the minimum value is $|E|$, this is necessarily less than $1/2$. One therefore has at least two points where $M=1/2$, one of which is the apparent horizon
for the black hole associated with the singularity at $z_S$. Therefore, as in the $D=0$ case, there will always be a black hole apparent horizon but  not necessarily
an event horizon. This emphasizes an important difference between the collapse singularities at $z=1/D$ and $z=z_S$: only the latter is associated with an apparent
horizon, which is why only the former can be naked.

The $D<0$ solutions have the same form as in the $D=0$ case, except that 
the $E<0$ ones have a second singularity at the value $z_S$ given by eqn (3.15) with $D<0$. In this case, as $z$ goes from
$-1/D$ to $z_S$, $S$ first increases to some maximum value and then decreases, while $V$ monotonically increases from $-\infty$ to $+\infty$.
As in the
$E=0$ case, such models are probably physically unrealistic since the density diverges due to shell-crossing. They are therefore not shown explicitly.
The form of the (unphysical) negative-mass solutions, which only exist for $E>0$, is indicated by the dotted curve in Figures (3). This is similar to the $D=0$
case shown in Figures (2) except that asymptotically eqn (3.19) applies rather than eqn (3.8).

\setcounter{equation}{0}
\section{Conclusion}

We may briefly summarize the results of our analysis as follows. (1) There are two families of spherically symmetric self-similar dust models: asymptotically flat
Friedmann solutions and what we have termed asymptotically quasi-static solutions. These all represent inhomogeneous cosmological models in which the energy function
$E$ is constant. They either expand from a Big
Bang or collapse to a Big Crunch but the singularity is only at $t=0$ for the asymptotically Friedmann family. (2) Some of the asymptotically Friedmann 
models
represent overdensities in a Friedmann background which recollapse to a second singularity and contain a black hole which grows as fast as the Universe. The black
hole always has an apparent horizon but not necessarily an event horizon. Other asymptotically Friedmann models represent underdensities in a Friedmann background
which grow as fast as the Universe.  (3) The asymptotically quasi-static models can be interpreted as representing either inhomogenenous cosmological
solutions (with one or two singularities) if one allows both signs of $z$ or self-similar collapse from an initially isothermal distibution if one allows just one
sign of
$z$, with a uniform density core in one regime and a central black hole or naked singularity in the other. (4) We
have emphasized the relationship between the
$z>0$ and $z<0$ solutions. Any particular asymptotically Friedmann solution is confined to one sign of $z$ but any asymptotically quasi-static solution necessarily
spans both signs. 

In an accompanying paper (Carr \& Coley 1999a), it is shown that the spherically symmetric self-similar solutions with pressure share many of
the  qualitative features of the dust ones, especially in the 
supersonic regime. In particular, all of the properties (1) to (4) above still pertain. However, it should be emphasized 
that new types of solution arise when there is pressure. For example, there is an exact static solution and an exact
Kantowski-Sachs solution, as well as families of solutions asymptotic to these. There are also asymptotically Minkowski solutions for $\alpha
>1/5$, some of which asymptote to a finite value of $z$. The inclusion of pressure obviously introduces qualitatively new features in the subsonic regime, in
particular the possible presence of a sonic point.

In claiming that our classification is ``complete'', it should be emphasized
that our considerations have been confined to similarity solutions of the simplest kind (i.e. homothetic solutions in which the similarity variable is 
$z\equiv r/t$). However, it should be noted that this is not the only type of similarity. For example, Carter \& Henriksen (1989) have 
generalized the concept to include what they term ``kinematic" self-similarity. In this context the similarity variable is of the form
$z=r/t^a$ for $a\neq 1$ and the solution may contain some dimensional 
constant. Ponce de Leon (1993) has also introduced the closely related notion of ``partial homothety". 
It is not yet
clear how easily the analysis of this paper can be extended to these cases. Finally it should also be emphasized that we have only been studying solutions
which are homothetic {\it everywhere}. The sort of models considered by Tomita (1997), in which one patches a self-similar transition region between other 
non-self-similar regions, is clearly not covered here. 

\vskip .3in

{\large \bf Acknowledgments}

The author thanks Alan Coley for useful discussions and is grateful to the Yukawa Institute for Theoretical Physics at Kyoto University and the Department of
Mathematics and Statistics at Dalhousie University for hospitality received during this work.

\newpage

\noindent
{\large \bf References}

 \begin{enumerate}

\item[] H. Bondi, 1947, MNRAS {\bf 107}, 410.

\item[] W. B. Bonnor, 1956, J. Astrophys. {\bf 39}, 143.

\item[] A. H. Cahill and M. E. Taub, 1971, Comm. Math. Phys. {\bf 21}, 1.

\item[] B. J. Carr and A. A. Coley, 1999a, Phys. Rev. D., in press.

\item[] B. J. Carr and A. A. Coley, 1999b, Class. Quant. Grav., {\bf 16}, R31.

\item[] B. J. Carr and S. W. Hawking, 1974, MNRAS {\bf 168}, 399.

\item[] B. J.  Carr and A. Yahil, 1990, Ap. J. {\bf 360}, 330.

\item[] B. J.  Carr and A. Whinnett, 1999, preprint.

\item[] B. Carter and R. N. Henriksen, 1989, Ann. Phys. Supp. {\bf 14}, 47.

\item[] D. Christodoulou, 1984, Commun. Math. Phys. {\bf 93}, 171.

\item[] I. H. Dwivedi and P. S. Joshi, 1997, Class. Quant. Grav. {\bf 47}, 5357.

\item[] C. C. Dyer, 1979, MNRAS {\bf 189}, 189.

\item[] D. M. Eardley and L. Smarr, 1979, Phys. Rev. D. {\bf 19}, 2239.

\item[] V. T. Gurovich, 1967, Sov. Phys.-Dokl. {\bf 11}, 569.

\item[] S. W. Hawking and G. F. R. Ellis, 1973, The Large-Scale Structure of Spacetime, Cambridge University Press.

\item[] R. N. Henriksen and K. Patel, 1991, Phys. Red. D. {\bf 42}, 1068.

\item[] P. S. Joshi and I. H. Dwivedi, 1993, Phys. Rev. D. 47, 5357.

\item[] P. S. Joshi and T. P. Singh, 1995, Phys. Rev. D. {\bf 51}, 6778.
 
\item[] R. Kantowski and R. Sachs, 1966, J. Math. Phys. {\bf 7}, 443.

\item[] A. Krasinzki, 1997, Physics in an Inhomogeneous Universe,  Cambridge University Press.

\item[] K. Lake, 1992, Phys. Rev. Lett. {\bf 68}, 3129.

\item[] R. Larson, 1969, MNRAS {\bf 145}, 271.

\item[] S. D. Maharaj, 1988, J. Math. Phys. {\bf 29}, 1443.

\item[] M. V. Penston, 1969, MNRAS {\bf 144}, 449.

\item[] A. Ori and T. Piran, 1990, Phys. Rev. D. {\bf 42}, 1068.

\item[] J. Ponce de Leon, 1993, Gen. Rel. Grav. {\bf 25}, 865.

\item[] A. M. Sintes, 1996, PhD thesis (University of Balearic Islands).

\item[] R. C. Tolman, 1934, Proc. Nat. Acad. Sci. USA {\bf 20}, 169.

\item[] K. Tomita, 1995, Ap. J. {\bf 451}, 1.

\item[] K. Tomita, 1997a, Phys. Rev. D. {\bf 56}, 3341.

\item[] K. Tomita, 1997b, Gen. Rel. Grav. {\bf 13}, 625.

\item[] Z. C. Wu, 1981, Gen. Rel. Grav. {\bf 13}, 625.

\end{enumerate}

\vskip .5in

\noindent
{\large \bf Figures}

FIGURE (1). This shows the form of (a) the scale factor $S(z)$, (b) the velocity function $V(z)$ and (c) the density function $\mu t^2(z)$ for the asymptotically
Friedmann dust models, the arrows indicating the direction of increasing time. These are described by 
a single parameter $E$ where $E=0$ in the exact Friedmann case: the $z>0$ solutions are overdense and collapse to black holes for $E<0$ (with an event horizon 
for $E>E_*$ since $V_{min}<1$) but they are underdense and expand forever for $E>0$. The $z<0$ solutions 
are just the time reverse of these. The dotted curve corresponds to a solution with negative mass; it is probably unphysical but relates to the
Kantowski-Sachs solution which arises when there is pressure. 

FIGURE (2). This shows the form of (a) the scale factor $S(z)$, (b) the velocity function
$V(z)$ and (c) the density function $\mu t^2(z)$ for dust models with $E=0$. Two different values of $D$ are shown in (a) and (b) but only one in (c). These solutions
necessarily span both positive and negative  values of $z$. For $D>0$ they represent monotonically expanding (solid) or collapsing (broken) solutions and the latter
contain a naked singularity ($V_{min}<1$) if $D$ exceeds some value $D_+$ (as assumed here). The $D<0$ models (dotted) undergo shell-crossing before encountering the
singularity and are probably unphysical since $V$ and $\mu$ go negative.

FIGURE (3). This shows the form of (a) the scale factor $S(z)$, (b) the velocity function
$V(z)$ and (c) the density function $\mu t^2(z)$ for the asymptotically quasi-static
dust solutions. These are described by two parameters ($D$ and $E$) but we assume that $D$ is fixed and not all the solutions in (a) and (b) are shown in (c). 
For $E>0$
the solutions resemble those in Figure (2), with both monotonically expanding (solid) and collapsing (broken) solutions. The collapse singularity is naked
($V_{min}<1$) if $E$ is less than a value $E_+(D)$. For $E<0$ there are also solutions which recollapse to a black hole (solid) or emerge from a white hole (broken),
as in the asymptotically  Friedmann case. As $E$ decreases, the last solution is the symmetrical one for which $E=E_{sym}$ and $z_S=1/D$, so that the solid and broken
curves coincide. The curves labelled $E=0^+$ and $E=0^-$ show the qualitative transition as $E$ passes through $0$. The
dotted curves correspond to an (unphysical) negative mass solution.

FIGURE (4). This shows the permitted regime for the parameters $E$ and $D$. The curve labelled $E_{sym}$ indicates the symmetric solution and all physical
solutions must lie above this. The upper broken line gives the transition between different asymptotic forms for $S_\infty$ and $V_\infty$. The collapsing solutions
have a naked singularity in the vertically shaded region below the line labelled
$E_+$ and the black hole solutions have an event horizon and a particle horizon in the horizontally shaded region above the line labelled $E_*$. 
These lines intersect on the lower boundary.

\end{document}